\documentclass[iop]{emulateapj}
\usepackage{graphicx}
\usepackage[suffix=]{epstopdf}
\usepackage{natbib}
\usepackage{amsmath}

\usepackage{hyperref}

\submitted{\today}
\journalinfo{}

\shortauthors{Davenport et al.}
\shorttitle{Gender in AAS Talks}

\begin{document}

\title{Studying Gender in Conference Talks -- data from the 223rd meeting\\
 of the American Astronomical Society}

\author{
	James R. A. Davenport\altaffilmark{1,2},
	Morgan Fouesneau\altaffilmark{1},
	Erin Grand\altaffilmark{3},
	Alex Hagen\altaffilmark{4},\\
	Katja Poppenhaeger\altaffilmark{5},
	and Laura L. Watkins\altaffilmark{6}
	}

\altaffiltext{1}{Department of Astronomy, University of Washington, Box 351580, Seattle, WA 98195, USA}
\altaffiltext{2}{Corresponding Author: jrad@astro.washington.edu}
\altaffiltext{3}{Department of Astronomy, University of Maryland, College Park, MD 20742-2421, USA}
\altaffiltext{4}{Department of Astronomy and Astrophysics, The Pennsylvania State University, 525 Davey Lab, University Park, PA 16802, USA}
\altaffiltext{5}{Harvard-Smithsonian Center for Astrophysics, Cambridge, MA 02138, USA}
\altaffiltext{6}{Space Telescope Science Institute, 3700 San Martin Drive, Baltimore, MD 21218, USA}

\begin{abstract}
We present a study on the gender balance, in speakers and attendees, at the recent major astronomical conference, the American Astronomical Society meeting 223, in Washington, DC. We conducted an informal survey, yielding over 300 responses by volunteers at the meeting. Each response included gender data about a single talk given at the meeting, recording the gender of the speaker and all question-askers. In total, 225 individual AAS talks were sampled. We analyze basic statistical properties of this sample. We find that the gender ratio of the speakers closely matched the gender ratio of the conference attendees. The audience asked an average of 2.8 questions per talk. Talks given by women had a slightly higher number of questions asked (3.2$\pm$0.2) than talks given by men (2.6$\pm$0.1).
 The most significant result from this study is that while the gender ratio of speakers very closely mirrors that of conference attendees, women are under-represented in the question-asker category.  We interpret this to be an age-effect, as senior scientists may be more likely to ask questions, and are more commonly men. A strong dependence on the gender of session chairs is found, whereby women ask disproportionately fewer questions in sessions chaired by men. While our results point to laudable progress in gender-balanced speaker selection, we believe future surveys of this kind would help ensure that collaboration at such meetings is as inclusive as possible.
\end{abstract}

\section{Introduction}
All scientific gatherings, from conferences to colloquia, should seek to be welcoming venues for intellectual exchange. Such meetings are, along with publications, the primary means by which scientists communicate. As we seek to improve the diversity of the body scientific, making our field more approachable to women, minorities, and other traditionally underrepresented peoples, we must ensure that our meetings also foster interaction between all members.

Anecdotal observations during a recent astronomy meeting noted a difference in the gender distribution between conference attendees, invited speakers, and the attendees that asked questions. More specifically, while the conference as a whole seemed well balanced in gender, as were the invited speakers, the questions appeared to be preferentially asked by men. Specifically, it appeared that men were asking the majority of the questions for every talk. We wondered the significance of this observation, and sought to gather more data on the subject.

Here we present findings from a semi-formal survey of oral presentations at a recent major astronomical conference, the 223rd Meeting of the American Astronomical Society (AAS), held in January 2014. The survey was a volunteer effort throughout the meeting, with submissions coming from anonymous attendees. The analysis was conducted as part of the AAS ``Hack Day'' program, which provided an excellent forum for discussion and creative input on the project. We hope this informal report will encourage further discussion and study of diversity and gender equality within our community.

\section{Collecting Data}
To gather data on the gender ratio of speakers and questioners we relied on a volunteer community effort. Our goal was to record the gender\footnote{We note that by using a binary gender distribution as classified by volunteers we are making an assumption that everyone is cisgendered and identifies as either male or female. We recognize that this assumption ignores valued members of the astronomical community who are transgendered and/or don't identify with a binary gender. Given the confines of this small study, we saw no way around this, but welcome suggestions about how to record and improve the recognition of scientists who are not represented by a binary gender.} of speakers and questioners from as many sessions and sub-fields as possible. 

In the weeks leading up to the AAS meeting we began advertising the survey. A link to the form was shared with colleagues using social media (Facebook and Twitter), as well as email, and posting on the popular professional astronomy blog \href{http://www.astrobetter.com}{\bf{www.astrobetter.com}} and the data visualization blog \href{http://www.ifweassume.com}{\bf{www.ifweassume.com}}. Tweets and Facebook posts were sent throughout the meeting to encourage continued participation by meeting attendees. 

The form consisted simply of three fields: 
\begin{itemize}
\item AAS Talk ID (e.g. 123.45)
\item Gender of the Speaker (Female or Male)
\item Gender of every Questioner (Female or Male)
\end{itemize}
The AAS Talk ID identifies both the session number (e.g. session 123) and order of the talks (e.g. talk 45). We asked respondents to record the gender of all questioners in the form MFMF, which would indicate a male question asker, followed by a female, then a male, and so on.

Concern was raised prior to the meeting about the use of binary gender identification, specifically that we may misrepresent or otherwise disenfranchise a subset of respondents or the meeting attendees under study.  In the interest of simplicity, we opted to maintain binary gender choices and invited feedback from anyone who found that this choice limited their participation in our study. No feedback on this point was received during the meeting, but we encourage members of the community who feel strongly about this point to reach out in the interest of future studies.

In total, we received 304 individual survey responses. Of these, 79 were duplicate entries where multiple responses were submitted for a single talk. We observed that many of these duplicates were the result of premature submission by a user, followed rapidly by an updated response. Several users also personally alerted us to their submission of duplicate responses. Thus we opted to use the ``longer'' entry (higher number of questions recorded) in the case of duplicates. Our final sample consisted of 225 talks from 53 individual sessions.


\begin{figure}[!t]
\centering
\includegraphics[width=3.25in]{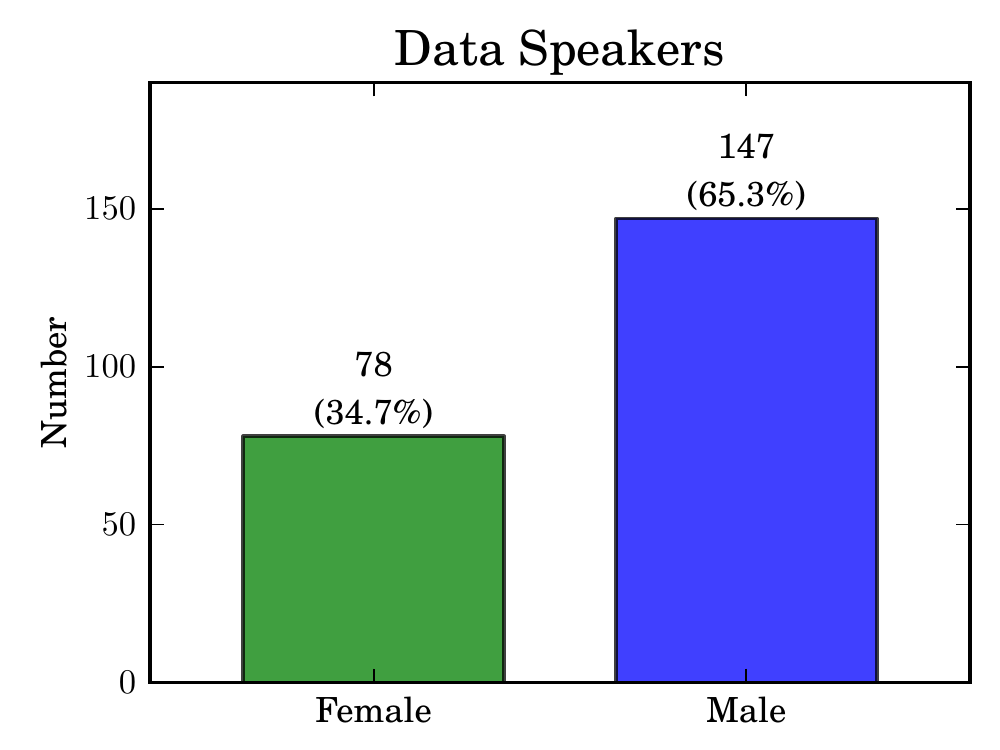}
\includegraphics[width=3.25in]{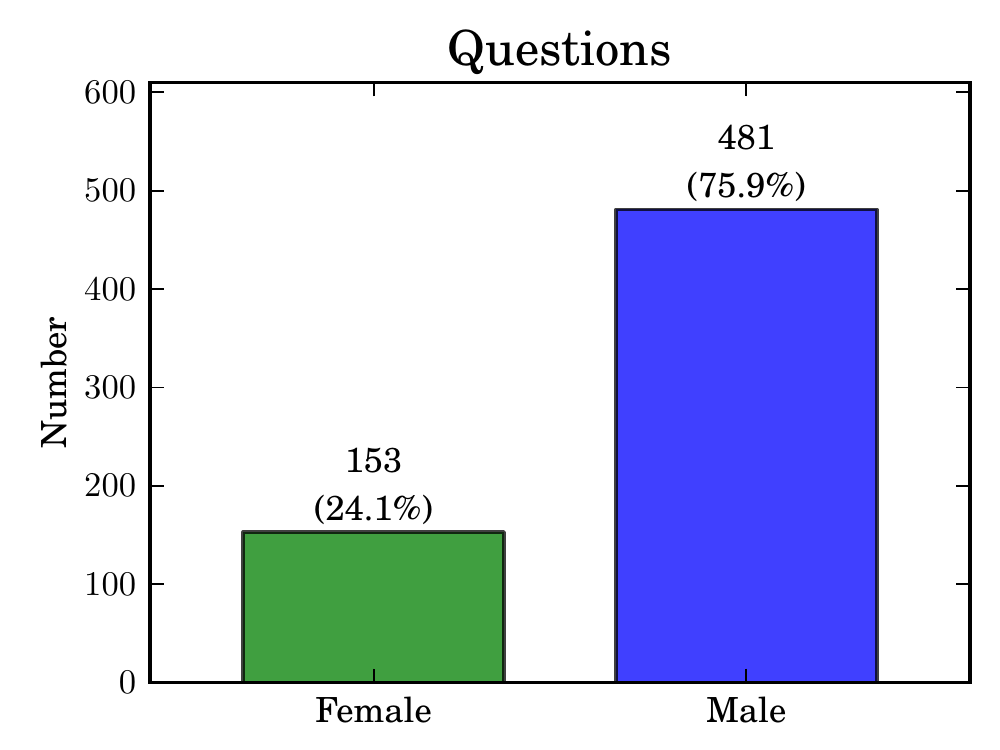}
\caption{Gender distribution for all speakers (top) and questioners (bottom) from our final sample.}
\label{fig:q}
\end{figure}

\begin{figure}[!t]
\centering
\includegraphics[width=3.5in]{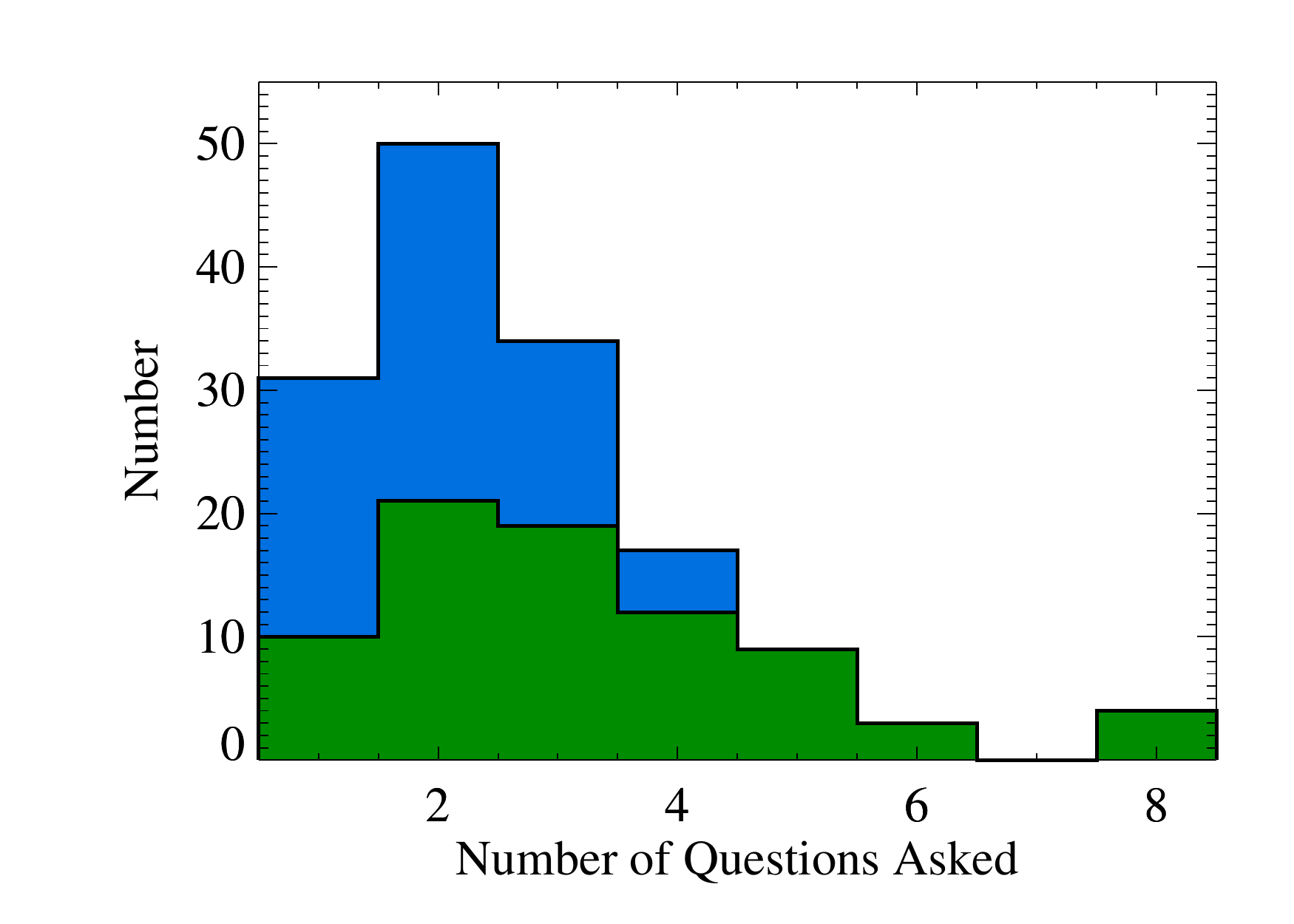}
\caption{Number of questions asked per talk as a function of the {\it speaker's} gender.}
\label{fig:qhist}
\end{figure}

\section{Basic Sample Properties}
Here we outline the general findings from our study of speakers and questioners, broken down by gender.  
Of the 225 individual talks in our sample, 78 (34.7\%) were given by women speakers, and 147 (65.3\% by men speakers. A total of 634 questions were recorded in our final survey sample. Of these, 153 (24.1\%) were asked by women, 481 (75.9\%) were asked by men. These results are shown in Figure \ref{fig:q}.

Intriguingly, for the 43 talks (19.1\%) with only a single question was asked, the questioner's gender ratio was closer to that of the overall speakers. For these talks we found 29 questions asked by men (67\%) and 14 by women (34\%). The robustness of this result is naturally to be questioned due to the small sample sizes. However, we speculate it may point to a different mode of speaker--questioner interaction, such as one questioner holding the floor, or a speaker going over their allotted time and only allowing for one question. Further study on this point is needed.

In Figure \ref{fig:qhist}, we show the distribution in the number of questions asked per talk as a function of the speaker's gender. For all talks, regardless of gender, the mean number of questions asked was 2.87. This number likely is strongly influenced by the time limits imposed for each presentation. We found the mean number of questions per talk for women speakers was slightly {\it higher} than for men, with averages of $3.28 \pm 0.20$ and $2.64\pm0.12$ respectively. Uncertainties quoted here are the standard error, defined as $\sigma/\sqrt{N}$ where $\sigma$ is the standard deviation in the number of questions per talk, and $N$ the total number of talks sampled. The distributions, however, are not gaussian, and so this difference is perhaps only a 1-$\sigma$ result. The underlying cause of such a difference in the number of questions asked is unclear at the moment, and a more stringent study is needed.

\begin{figure}[!th]
\centering
\includegraphics[width=3.25in]{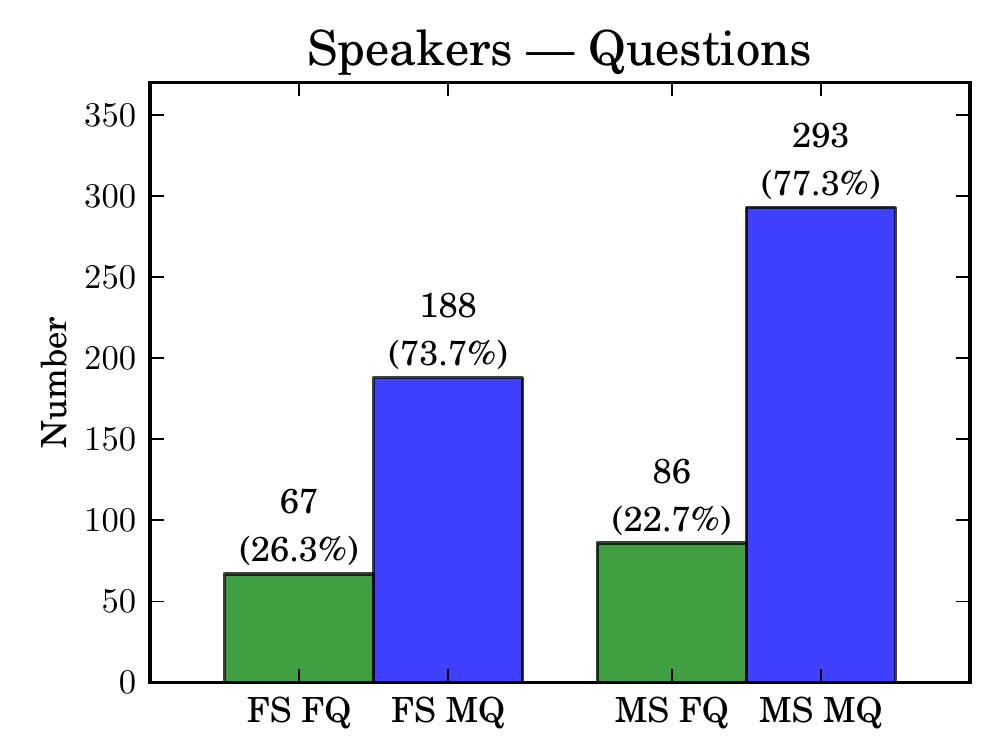}
\caption{The gender ratios between questioner and speaker genders. ``FS FQ'' indicates Female Speaker and Female Questions, ``FS MQ'' indicates Female Speaker and Male Questions, and so on}
\label{fig:sq}
\end{figure}

Comparing gender distributions of questions as a function of speaker gender showed only marginal trends. In Figure \ref{fig:sq}, we present our data broken down by both speaker and questioner gender. The difference in these distributions based on speaker gender is minimal.  If the gender of the speaker greatly affected the audience's tendency to ask question, as one might think is possible, a stronger difference between the male and female speaker ratios would be expected.

\clearpage

\section{Comparing to the Conference Attendees} 
To give the gender ratios found in the previous section meaning, we must place them in the context of conference attendees as a whole. To first order, we would expect that the gender ratios for speakers and questioners for such a large conference should be reflective of the conference attendee population. Determining whether either of the speaker or questioner gender distributions is reflective of the entire field of astronomy is beyond the scope of this work, though we encourage the AAS and others to investigate this point.

No easily parseable version of the conference program was available by the AAS website. 
On the final day of the conference, we downloaded the PDF version of the program, and with considerable labor extracted the names of all speakers. We found a total of 872 individual talks were scheduled for AAS 223, from 153 separate sessions.\footnote{We considered only talks from plenary and splinter sessions. Analysis of panel discussions is beyond the scope of this survey.} Our survey response data therefore sampled 25.8\% of all talks.

The AAS provided a very convenient listing of all conference attendees online, listing both their given names and surnames. We retrieved this listing of conference registrants on the first day of AAS 223, finding 3132 individuals listed online. As gender was not recorded for each registrant, we used the given (first) names to produce likely gender assignments.

To estimate genders based on first name, we used a unique dataset provided online\footnote{http://www.ssa.gov/OACT/babynames/limits.html}  by the United States Social Security Administration (SSA). For every year of birth, ranging from 1879 through the present, the SSA provides every unique first name for people who applied for Social Security Numbers, as well as the numbers of men and women with that name who applied in each year. This has provided valuable insight in to the trends of given names and genders over the past century.

We used every year of birth from 1960 onward to avoid any dramatic changes in gender--name conventions over this long timespan. We computed the total number of men and women who were given each unique first name in this dataset, and then divided each by the total number of people with that name. Gender for any given name could then naively be assigned using these ratios. We assumed a binary, majority takes all classification of name--gender assignment, such that if the majority of people in the SSA data with a given name were female, we assumed that name  was female.

We then matched the first names of AAS registrants to this SSA dataset, using literal string matching to lower-case versions of both lists, and ignoring character accents. From the list of 3132 registrants, 92.9\% were successfully matched to the SSA dataset. Of these,  34.0\% were determined to be women, and 66.0\% men. This gender ratio very closely matches that of the speakers found in \S3 above.

\begin{figure}[!t]
\centering
\includegraphics[width=3.5in]{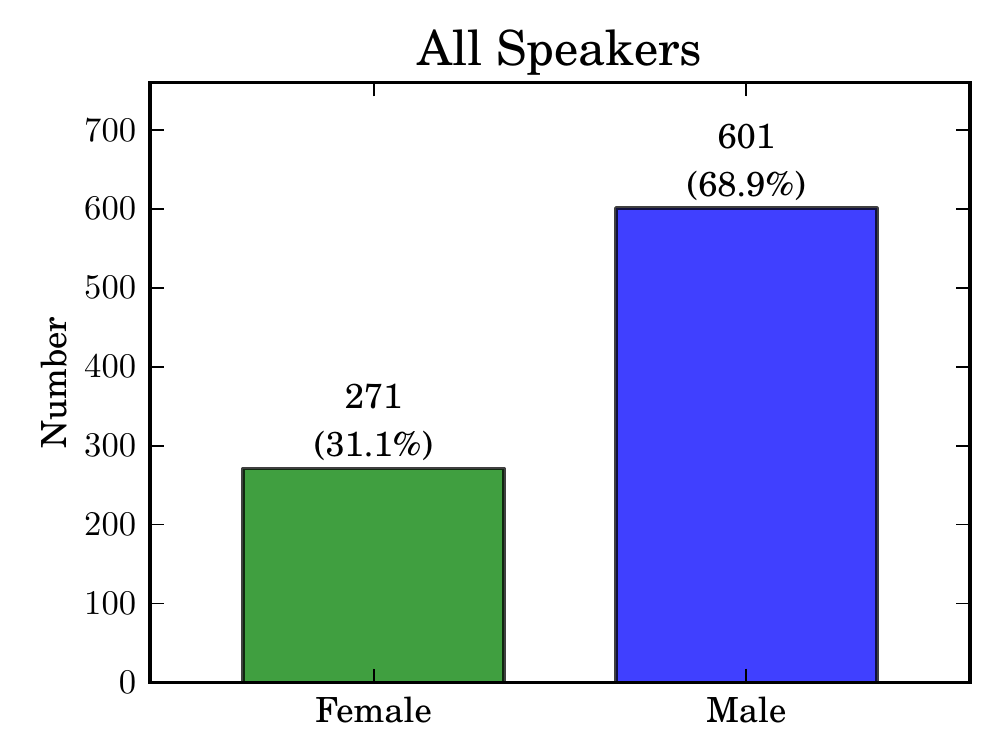}
\includegraphics[width=3.5in]{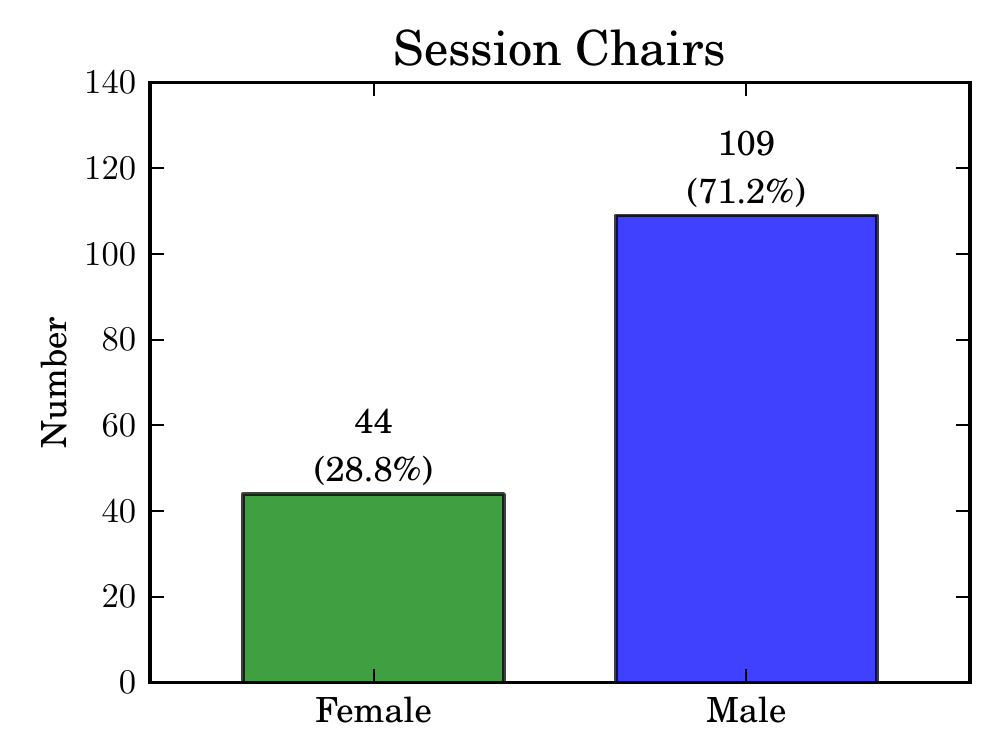}
\caption{Gender ratio for all conference speakers (top) and session chairs (bottom) listed in the AAS 223 program.}
\label{fig:speakerchair}
\end{figure}

\section{Effect of Session Chair Gender}

As an ancillary data product from our parsing of the AAS 223 meeting program, we gathered a list of the names for all session chairs as well as all speakers. Genders for these individuals were assigned in the same fashion as for conference attendees described above. These results are shown in Figure \ref{fig:speakerchair}. Broadly speaking, both the total speaker and session chair gender ratios followed that of the total conference attendees. Importantly, the gender ratio for all speakers was very close to the ratio for speakers we gathered data on, indicating our sample was drawn from a fairly unbiased subset of talks.

\begin{figure}[!t]
\centering
\includegraphics[width=3.5in]{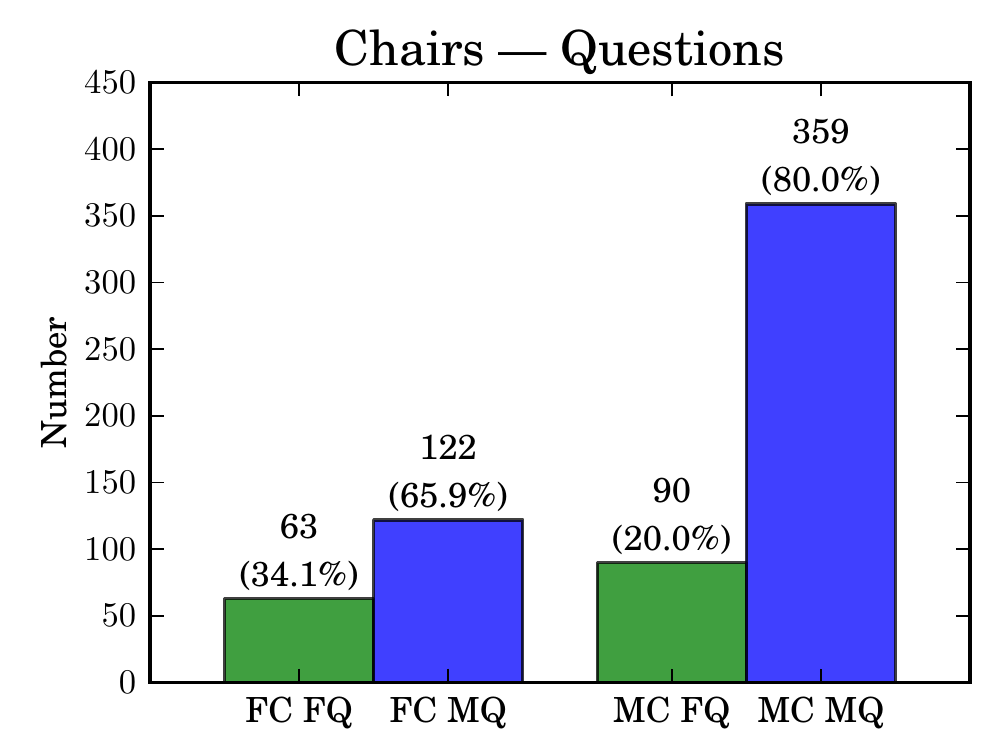}
\caption{Gender ratio of questioners as a function of speaker gender for the sessions our survey gathered at least 1 talk from. The bars are labeled with chair- and question-gender pairs (e.g FC FQ = Female Chair + Female Questions, and so on.)}
\label{fig:chairs}
\end{figure}

We then investigated the effect of session chair gender on the questions. In Figure \ref{fig:chairs}, we present the questioner gender ratios split on the session chair genders, similar to Figure \ref{fig:sq}. A significant dependence on session chair gender is found. Stunningly, while sessions chaired by women  have gender ratios of questions that very closely match the gender distribution of conference attendees, sessions with male chairs do not. Instead, these sessions disproportionately receive male questions with even greater skew than the results shown in Figure \ref{fig:sq} for Male Speakers. 

Speaker gender seems to have the greatest impact on the gender ratio of questioners. Our dataset is too limited to determine the proximate cause of this observation, and there exist many possible explanations. To give just one example, session chairs themselves may be asking questions. A more detailed study that accounts for the involvement of session chairs will be necessary.


\section{Reflection on Methodology}
The need for a more controlled and better organized study is evident. Users reported confusion about the layout of our web-form, as well as needing a more unambiguous way to record a lack of questions asked in a talk. Several users also reported not preserving the specific gender-order of questions asked, instead summing up all men and women questioners at the end of each talk.

Prof. David Hogg (NYU) suggested we include a feedback or comment box to capture metadata. This could be used to provide valuable user-reported insights in to, for example, the engagement of session chairs or the style/format of the Q/A sessions. 
We also received limited data from panel-style discussions and press releases, but were unsure how to incorporate it in our study.

\section{Concluding Remarks}

The primary results of this study are as follows:
\begin{enumerate}

\item Men ask disproportionally more questions than women in talks, despite the gender ratio of the speakers matching that of the conference attendees.\vspace{0.04in}

\item Women are asked slightly more questions per talk than men (3.28 versus 2.64, respectively.)

\item The gender of the session chair appears to have a strong correlation with the gender ratio of the questioners.

\end{enumerate}

We believe the first result may be explained by a simple line of reasoning: in the past the gender distribution of astronomers was very lopsided (strongly male). More senior scientists may simply be more likely to ask questions as deep experts. Speakers are instead drawn from a younger sample of scientists who are more often advertising new and novel work. This is supported by the recent Demographics Survey of 2013 US AAS Members, which found that the female/male split for astronomers born before 1980 was 21\%/79\%, while those born after 1980 were 40\%/60\%. These numbers very closely match the gender distributions we show in Figure \ref{fig:q}.

Given the modest effort in advertisement and organization before the meeting for our survey, we are encouraged by the participation and coverage of AAS talks (26\%) sampled here. Further, introspective studies of our profession are clearly supported by our community. In the immediate future, we would like to have a larger scale study, with particular focus one the impact of session format and session chair involvement. We believe this could be conducted by the AAS with very low organizational overhead and volunteer labor. Our pilot study has shown the great value of volunteer submitted data, and with minimal advertisement we believe data could be gathered for nearly 100\% of the conference talks.

Near term, we seek to improve meeting organization, striving to find the right scenario for presenting scientific advances between peers in a manner that is most inclusive and inviting to underrepresented peoples in our field. A recent demographic survey conducted by the AAS indicates that, while several major hurdles for equality in our field still exist, gender ratios are flattening with time.

In the not-too long term, we hope that the advancement of scientific knowledge will be a truly inclusive endeavor which benefits from contributions of all groups of scientists.

\acknowledgements
We graciously thank all the volunteers who helped gather data and provided many useful suggestions throughout the project. Many thanks also to the organizers and sponsors of the AAS 223 Hack Day.

\end{document}